\documentstyle[aps,pra,epsfig]{revtex}

\begin{document}
\title{\large \bf Law of Malus and Photon-Photon Correlations:\\
  A Quasi-Deterministic Analyzer Model}
  \author{Bill Dalton\thanks{Email: bdalton@stcloudstate.edu}}
\address{Department of Physics, Astronomy and Engineering Science\\
St. Cloud State University\\
St. Cloud, MN 56301, USA}
\date{\today}
\maketitle
\begin{abstract}

For polarization experiments involving photon counting we
introduce a quasi-deterministic eigenstate transition model of the
analyzer process. Distributions accumulated one photon at a time,
provide a deterministic explanation for the law of Malus. We
combine this analyzer model with causal polarization coupling to
calculate photon-photon correlations, one photon pair at a time.
The calculated correlations exceed the Bell limits and show
excellent agreement with the measured correlations of [ A. Aspect,
P. Grangier and G. Rogers, Phys. Rev. Lett. 49 91 (1982)]. We
discuss why this model exceeds the Bell type limits.
\end{abstract}
%

\section{INTRODUCTION}

Macroscopic electromagnetic field theory leads directly to the law
of Malus for light polarization measurements. For experiments
where the distributions are accumulated from in-sequence photon
counting, does the law of Malus still hold, and if so, what gives
rise to it? Measurements using a thin polarizing film have
supported the Malus formula for in sequence photon counting
\cite{Papal}. In these experiments where photons are counted
individually in detectors, it is difficult to justify the Malus
distribution from macroscopic field arguments. Past efforts to
explain the law of Malus for in-sequence photon counting
experiments, as well as photon-photon correlations, with hidden
variable models have been unsuccessful
\cite{Belinf,Ballen,Afriat}. These models involve a strong
probability component.

Could we successfully explain the law of Malus and measured
correlations by replacing the probability feature by a
deterministic decision process, accumulating the distributions one
particle, or particle pair at a time? The affirmative answer to
this question is the focus of this paper. A deterministic
trajectory model \cite{Dal-G} has been previously used to explain
"ghost diffraction" patterns \cite{Strek}. Here, we introduce a
quasi-deterministic model of the analyser process. This model
involves an independent stochastic variable in each analyzer, and
a deterministic criterion for selecting the polarization
eigenchannel in the analyzer. In this model, photons are viewed as
field wave packets and trajectory calculations are not involved.

We use this model in detailed calculations to explain the law of
Malus for in-sequence photon counting. We then use this same model
to successfully explain some well known photon-photon correlations
measurements of \cite{Aspect-81,Aspect-82,Alley}. In
\cite{Dal2000} we use this quasi-deterministic analyzer model to
sucessfully explain the proton-proton correlations of \cite{LR-M},
as well as the four-angle photon-photon correlations of Aspect et
al. \cite{Aspect-82}. This model makes use of causal polarization
coupling between photons in a pair and local stochastic variables.
\section{STOKES REPRESENTATION OF THE EIGENVALUE PROBLEM}

For crystal beam splitters, beam separation is based on
eiegenstates, and in particular, the eigenvalues of the refractive
index matrix \cite{Yariv}. Because of the difference in the
eigenvalues for the \textit{e} and \textit{o} eigenstates, it is
possible to separate these states. Because of the dependence on
the eigenvalues, we call this type of splitter an eigenvalue
splitter. The quasi-deterministic model described in detail below
is based on eigenchannel selection. Stern-Gerlach type analyzers
for spinors are likewise eigenvalue splitters. In the model
described here, we introduce a particular stochastic variable in
the refractive index matrix. Changing this particular variable
does not affect the eigenvalues, so that the usual beam separation
arguments are unaffected.

Polarization is conveniently represented by Stokes variables and
rotations on the Poincare' sphere. Stokes representations for both
spinors and vectors are similar and well developed \cite{McMas}.
To formulate this eigenstate transition model, we first rewrite
certain relations contained in the generic two-component
eigenvalue equation in terms of a convenient Stokes variable
representation. All analysis here involves a two dimensional
complex representation for the field.

Consider a Hermitian matrix with elements $h_{11} = a$,
$h_{22}=d$, $h_{12}=h\exp(-\imath\phi)$, and
$h_{21}=h\exp(\imath\phi)$, where $a$, $d$, $h$, and $\phi$ are
real. With $\Psi^T= {\left[\Psi_x,\Psi_y\right]}$,
$\Psi_x=A\cos(\beta)\exp(\imath\alpha_x)$ and
$\Psi_y=A\sin(\beta)\exp(\imath(\alpha_x+\delta)$, we write the
matrix eigenvalue equation $h\Psi  = \lambda\Psi$ in component
form as follows.
\begin{equation}\label{aa1}
(a-\lambda)\cos(\beta)+h\exp(\imath(\delta-\phi))\sin(\beta)=0
\end{equation}
\begin{equation}\label{aa2}
h\exp(-\imath(\delta-\phi))cos(\beta)+(d-\lambda)\sin(\beta)=0
\end{equation}
 In terms of the above variables, the two eigenvalues are given by the following
 formula.
\begin{equation}\label{aa3}
\lambda_\pm = (a+d \pm \sqrt{(a-d)^2+4h^2})/2
\end{equation}
We employ two Stokes sets $S$ and $P$ that represent the field and
matrix respectively. In terms of the field variables above, the
components of $S$ have the familiar \cite{McMas} form $S_0 =
\Psi^\dagger\Psi$, $S_1 = S_0\cos(2\beta) $, $S_2 =
S_0\sin(2\beta)\cos(\delta)$, and $S_3
=S_0\sin(2\beta)\sin(\delta)$. The components of $P$ are given by
$P_0 = \sqrt{(a-d)^2+4h^2}$, $P_1 = P_0\cos(2\alpha)$, $P_2 =
P_0\sin(2\alpha)\cos(\phi)$ and $P_3 =P_0\sin(2\alpha)\sin(\phi)$,
where $\tan(2\alpha)=2h/(a-d)$.
 In terms of
these components of $S$ and $P$, we can extract from equations
(\ref{aa1}), (\ref{aa2}), and (\ref{aa3}) the following relations.
\begin{equation}\label{eig1}
\textbf{\textsl{P}}\cdot\textbf{\textsl{S}}=\pm P_0S_0
\end{equation}
\begin{equation}\label{eig2}
S_3P_2-S_2P_3=0
\end{equation}
\begin{equation}\label{eig3}
S_1P_1=\pm(P_1)^2S_0/P_0
\end{equation}
Equation (\ref{eig2}) means $\sin(\delta-\phi)=0$ and follows
directly from the imaginary component of (\ref{aa1}) and
(\ref{aa2}).

The vectors $\textbf{\textsl{S}}$ and  $\textbf{\textsl{P}}$
represent points on two Poincare' polarization spheres with common
centers but with different radii. The radius of the $S$ sphere is
given by $S_0 = \Psi^\dagger\Psi=A^2$. The radius of the $P$
sphere is given by  $P_0=\lambda_+ -\lambda_-$. The Stokes
variables rotate with twice the rotation angle of the field
components. If the space coordinates are rotated by $\theta$ say,
the Stokes variables for a spinor field are rotated by $\theta$
also, whereas the Stokes variables for a vector field are rotated
by $2\theta$.

In this model, an incident field makes a transition to an
eigenstate described by equations (\ref{eig1}), (\ref{eig2}), and
(\ref{eig3}).  From (\ref{eig3}) we see that the eigenvalue sign
choice in (\ref{aa3}) is related to the sign of the product
$S_1P_1$. We will make use of this relation in the deterministic
transition criteria discussed below.
\section{STOCHASTIC ANALYZER VARIABLE}

The traditional \textit{e} and \textit{o} rays of classical
macroscopic electro-optics correspond to two points on opposite
sides of the Poincare' sphere indicated by $S_1=\pm S_0$ where in
the diagonal frame we have $P_1=P_0$ for one crystal type. The two
eigenvalues, on which the separation decision is made, determine
the $P$ sphere radius via $\lambda_+ -\lambda_- = P_0$, but not
the direction of $\textbf{\textsl{P}}$. For a fixed relative phase
$\phi$, this gives one free variable for the matrix Stokes vector.

The point of view here is that the classical matrix Stokes vector
used for a macroscopic field of many photons does not necessarily
represent the matrix Stokes vector experienced by individual
photons. The first assumption of this model (The distributive
assumption) is that the $\textbf{\textsl{P}}$ vectors experienced
by individual incident pulses are distributed in the one free
variable, at least for the surface transition region. For a given
incident pulse, we randomly select this degree of freedom of
$\textbf{\textsl{P}}$ from a distribution described below.

There are two questions related to the depth $(z)$ dependence in
this model. First, are the residual deviations of $P_1(z)$ from
the macroscopic field value only surface effects, rapidly
decreasing with depth into the crystal, or, do they extend
throughout the crystal? Both cases give the same correlations, and
both cases give rise to the law of Malus. To correctly describe
the correlations, or the law of Malus, it is only necessary with
the transition criterion described below, that the surface value
$P_1(0)$ is selected from a distribution described below. For a
macroscopic field of many photons, one would not expect to have a
distributive $\textbf{\textsl{P}}$ vector because of averaging. It
is therefore doubtful that we can obtain information about this
distribution question with experiments using a macroscopic field
of many photons.

 Second, at what depth into the crystal does the
incident field change so that equations (\ref{eig1}),
(\ref{eig2}), and (\ref{eig3}) are satisfied? This
transition-depth problem in general has been the subject of
debated over many years \cite{Fearn}. Unfortunately, the analysis
here provides no answer to these questions. The presence of a
local stochastic variable should have some observable affect.
 Indeed, it does. Within this model, this stochastic variable makes it possible
to correctly describe features of both the law of Malus and
photon-photon correlations. With the distribution selected, the
average of $P_1(0)$ is the classical macroscopic matrix stokes
variable.
\section{STOCHASTIC VARIABLE SAMPLING}

For a frame attached to the crystal analyzer (say aligned with the
$\textit{e}$ and $\textit{o}$ rays), we can partition the matrix
Poincare' sphere with hemispheres indicated by the sign of $P_1$.
However, viewed from a frame attached to one analyzer, this
partition for the other analyzer is rotated. With respect to an
analyzers attached frame, we choose $\textbf{\textsl{P}}$ by
choosing $2\alpha$ via $2\alpha=arg$ where $arg =
\arccos(u)-\pi/2$ and $u$ is selected uniformly from the interval
$[-1,1]$. For our linear polarizer, we have $\cos(\phi)=\pm1$ in
the crystal frame. Viewed from a frame fixed to the first
analyzer, the $\textbf{\textsl{P}}'$ hemisphere axis for the
second analyzer is rotated from the first by $q=\theta$ for
spinors and $q=2\theta$ for vectors. In this fixed reference
frame, sampling for the second analyzer is made using
$2\alpha'=arg(u') \pm q$ where $u'$ is selected uniformly from the
interval $[-1,1]$, but independent of $u$.  The independence of
the selection of $u$ and $u'$ represents a local stochastic
element of this theory, and is the reason why we call this a
quasi-deterministic model.

One should ask if the above sampling distribution is the only
distribution that can successfully described the data. There are
many distributions with which one can exceed the Bell limits in
this model. The above distribution and a Gaussian distribution
with approximately the same width are the only two distributions
that the author has found to date that correctly describe the data
for both the law of Malus and the photon-photon correlations.
\section{DETERMINISTIC TRANSITION CRITERION}

The second assumption of this model (The deterministic assumption)
is that the incident pulse makes a transition to one eigen-channel
or the other, and that the choice is made with a deterministic
criteria based on initial conditions of the incident pulse at the
analyzer and the randomly selected $\textbf{\textsl{P}}$. From
(\ref{eig2}) we see that the sign of $S_1P_1$ indicates the
eigen-channel choice. We indicate this product as a function of
the depth $(z)$ into the crystal, as follows.
\begin{equation}\label{TRA}
T(z)=S_1(z)P_1(z)
\end{equation}
The deterministic criteria on which the calculations here are
based is that the sign of $T(z)$ after the transition is the same
as, and determined by the sign of $T(0)$. For a given value for
$S_1(0)$ and $P_1(0)$ we make the eigen-channel decision by
testing the sign of $T(0)$. In these calculations, we only use one
of the three equations (\ref{eig1}), (\ref{eig2}), and
(\ref{eig3}). Modeling transition criteria using all three
equation is a subject of ongoing study by the author.
\section{LAW OF MALUS}

\begin{figure}[htbp]
  \vspace*{19.00cm}
  \hspace*{5.0cm}
 \includegraphics{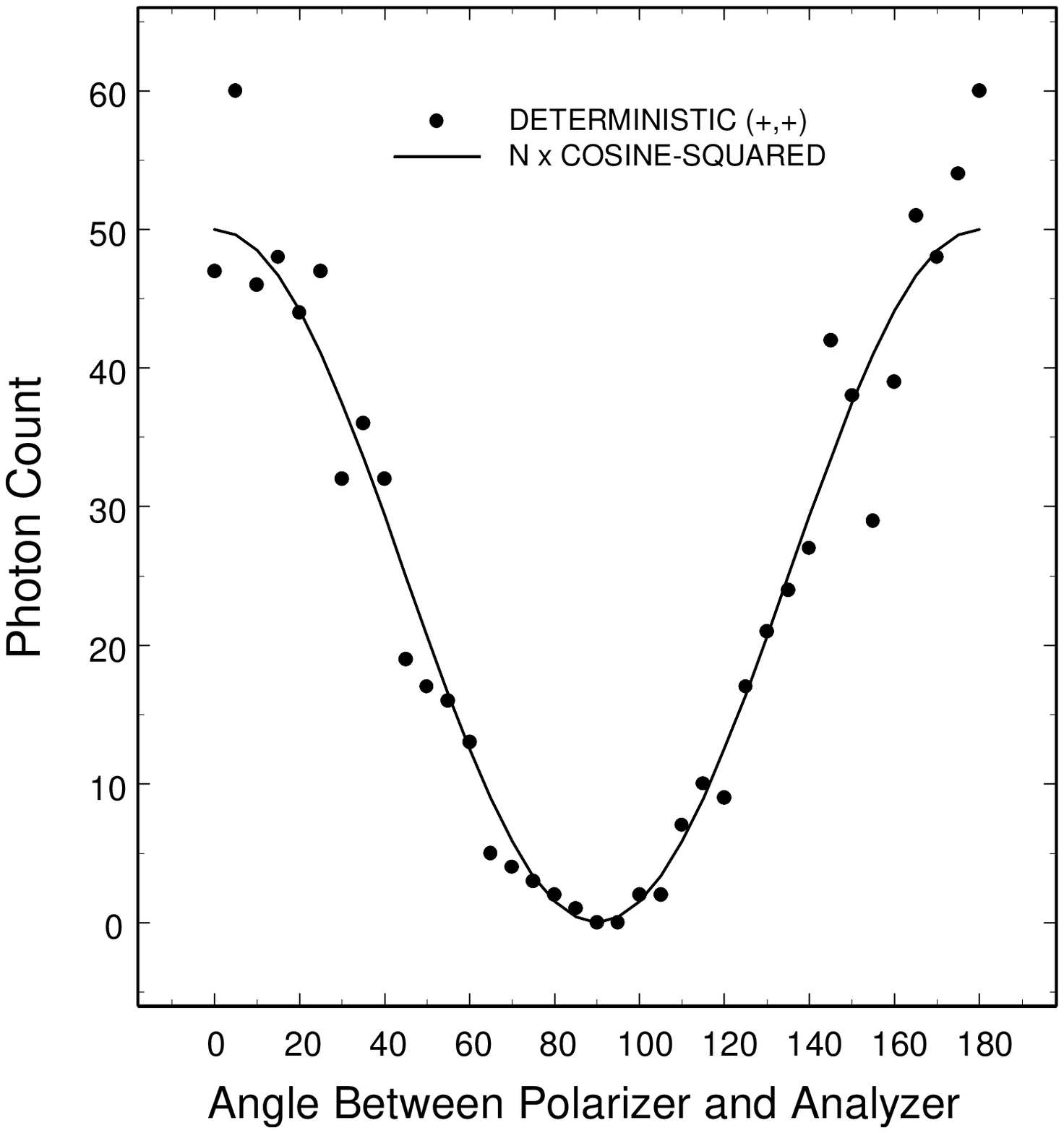}
 \vskip -1.3in
    \caption{Malus distribution for few photons. Calculated using the
    deterministic\\
   analyzer model: Solid circles indicate the photon
   accumulation.
    Counts are \\ accumulated one photon at a time.}
    \label{f1}
\end{figure}

\begin{figure}[htbp]
  \vspace*{19.00cm}
  \hspace*{5.00cm}
  \includegraphics{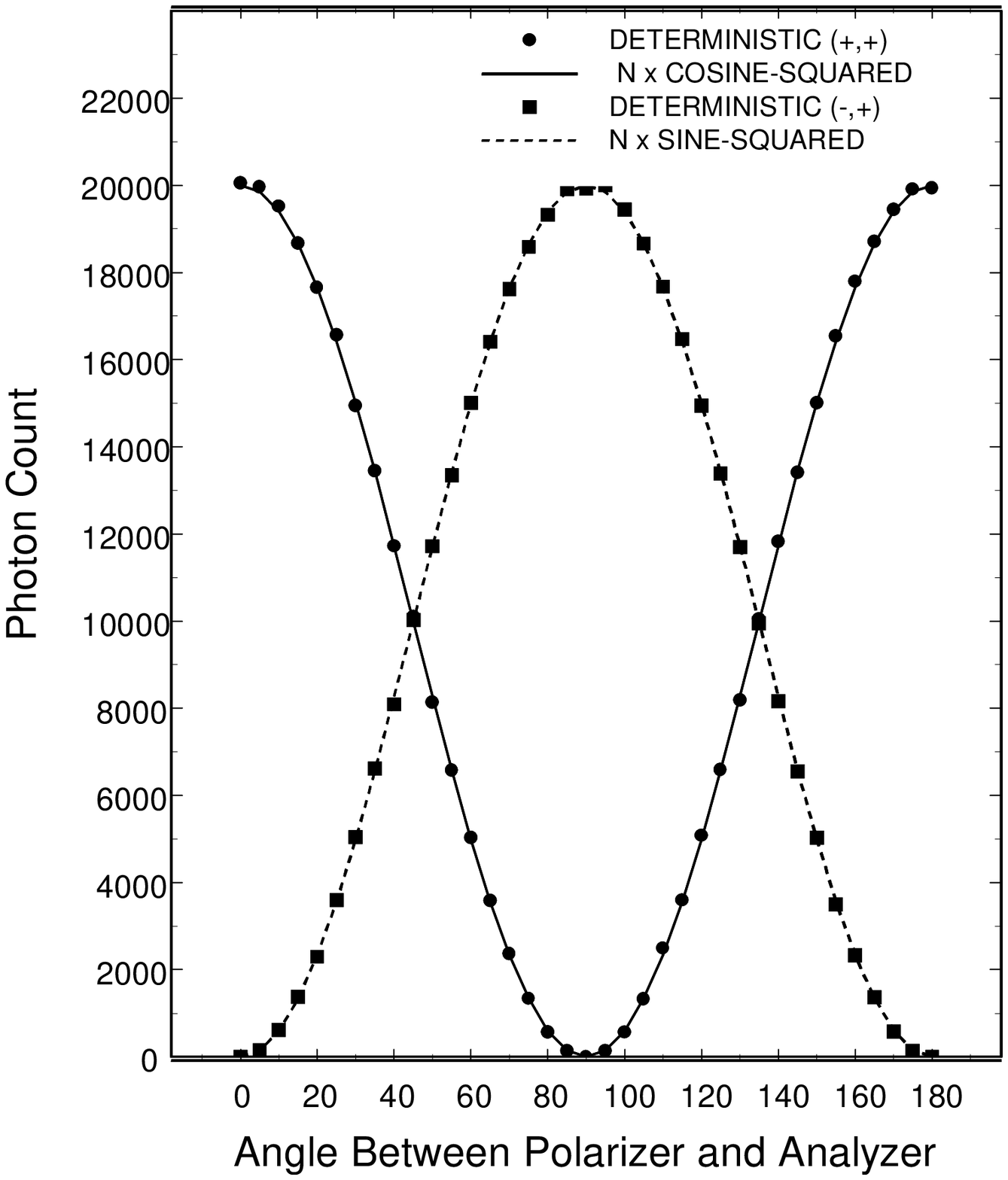}
 \vskip -1.0in
    \caption{Malus distribution for many photons. Calculated using the
    deterministic\\
   analyzer model: Solid circles indicate the photon
   accumulation.
    Counts are \\ accumulated one photon at a time.}
    \label{f2}
\end{figure}

In the study here, we use two deterministic analyzers in sequence,
with the second rotated at an angle $\theta$ relative to the
first.  We view the second analyzer in the reference frame of the
first.  The calculations are made one photon at a time. We use the
same incident number of photons at each relative angle. The
notation $(+,-)$ indicates the configuration for photons that exit
the $(+)$ channel of the first analyzers and the $(-)$ channel of
the second. To illustrate the statistical nature of the
distributions, we make calculations for both a small ($100$), and
a large ($40000$) number of incident photons. For the small number
case, the accumulated photon counts at each angle for the $(+,+)$
configuration are indicated by the solid circles in Fig.~\ref{f1}.
The fluctuations represent the approximate Poisson statistics of
the sampling. The output for the large number case for the $(+,+)$
configuration is indicated by the solid circles in Fig.~\ref{f2}.
 The solid line represents the scaled Malus formula
$N\cos^2(\theta)$ where $N$ is half the incident photons. The
solid squares in Fig.~\ref{f2} represent the accumulated
distribution for the $(-,+)$ case.  The dashed line represents the
formula $N\sin^2(\theta)$.  From these results, it is clear that
this quasi-deterministic hidden variable model clearly gives rise
to the law of Malus for in-sequence photon counting.
\section{PHOTON-PHOTON CORRELATIONS}

\begin{figure}[htbp]
  \vspace*{19.00cm}
  \hspace*{5.00cm}
  \includegraphics{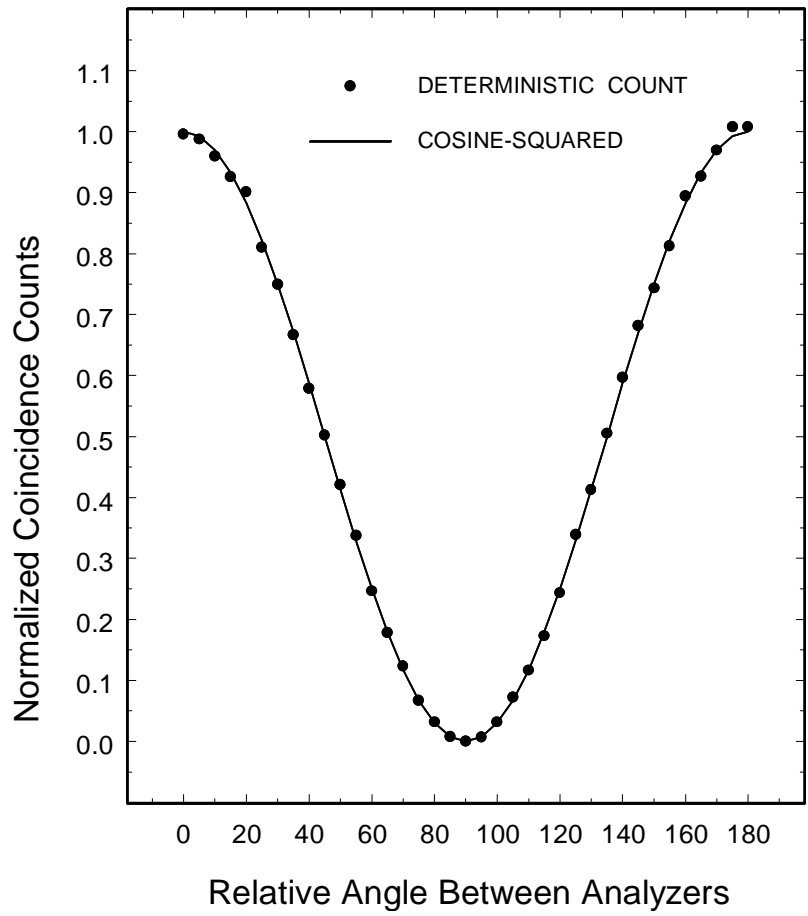}
 \vskip -1.0in
    \caption{ Normalized photon-photon coincidence counts $ 2N_{++}/N$
     versus relative\\
angle: These calculations used the quasi-deterministic analyzer
model. The \\ distributions are accumulated one photon pair at a
time.}
    \label{f3}
\end{figure}

To make calculations for photon-photon coincidence measurements,
we use two deterministic analyzer models to represent the two
receiving analyzers.  Pair counts are accumulated via
deterministic calculations, one photon pair at a time. The
parameters $u$ and $u'$ for the two analyzers are sampled as
indicated above, but independently of each other. Pair counts for
the four different coincidence combinations are represented here
by $N_{++}$ ,$N_{+-}$ , $N_{-+}$ and $N_{--}$ . For instance
$N_{+-}$ is the count for the number of pairs with a $(+)$ for the
first analyzer an $(-)$ for the second analyzer. The correlations
are calculated using the function $\gamma(\theta)=(N_{++}+
N_{--}-N_{+-}-N_{-+})/N$ \cite{Garucc}, where $ N =N_{++}+
N_{--}+N_{+-}+N_{-+}$ is the total number of pairs counted. For
the causal coupling, we used $S_{1}'(0)= S_1(0)$ where the
parameter $2\beta$ was chosen randomly from the interval
$[0,2\pi]$.  In the correlation calculations here, the deciding
information is in the common sign of the first factors in the
transition functions (\ref{TRA}) for each analyzer. It is only
necessary that $S_1(0)$ and $S_1'(0)$ have the same sign.
 In Fig.~\ref{f3} the solid circles
  represents $2N_{++}/N$. This curve follows the $\cos^2(\theta)$ formula,
  and agrees well with the
  coincidence measurements of \cite{Aspect-81,Aspect-82}, as well as the results
   reported in \cite{Alley} for the coincidence data obtained using a
   narrow spectral bandwidth and a thin crystal for the down-conversion process.

\begin{figure}[htbp]
  \vspace*{19.00cm}
  \hspace*{5.00cm}
  \includegraphics{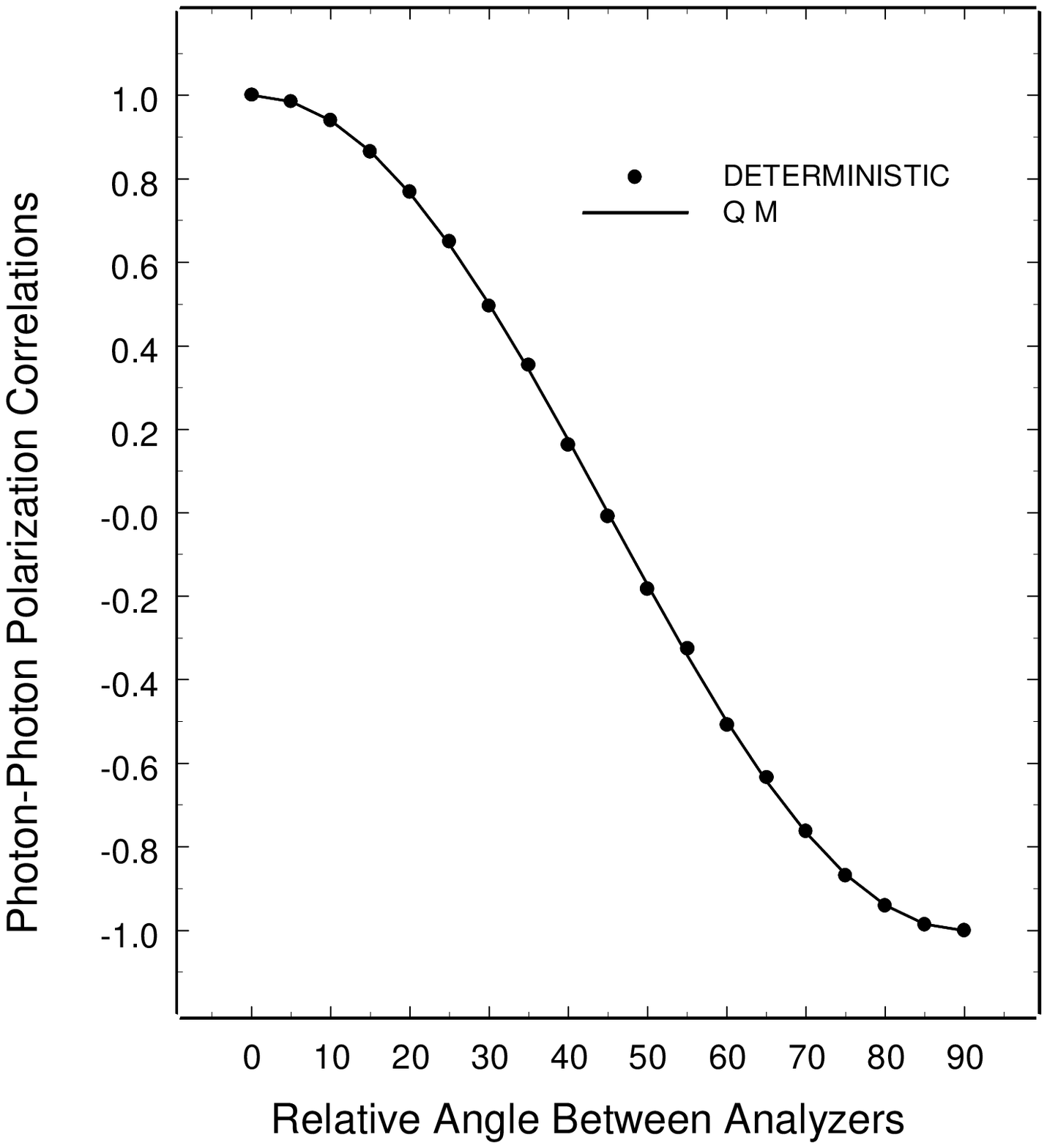}
 \vskip -1.0in
    \caption{Photon-photon correlations calculated using the
quasi-deterministic analyzer \\ model: Solid circles represent the
calculated correlation obtained via accumulation, \\ one photon
pair at a time. The line indicates the quantum mechanics
prediction.}
    \label{f4}
\end{figure}

The solid circles in Fig.~\ref{f4} represent the calculated
correlation function  \cite{Garucc,Aspect-82}. The four
pair-counts were accumulated, one photon pair at a time.
 Calculations are made using about $10000$ photon pairs
at each angle. The solid line in Fig.~\ref{f4} represents the
formula of quantum mechanics.  The two curves show excellent
agreement with each other and with the measurements of
\cite{Aspect-82}. We emphasize that all contributions to
$\gamma(\theta)$ were calculated using the four pair counts only
so there is no chance that an accidental scale factor can be
included.
\section{EXCEEDING BELL'S LIMITS}

The causal polarization link between the two photons in a pair
produces an angular correlation link even with the presence of the
local stochastic variables in the analyzers. This angular
dependent link is why the usual inequality Bell's theorem
\cite{Bell,CHSH} doesn't apply to this quasi-deterministic
analyzer model.

To explain this angular dependence in more detail, consider the
transition function
$T(0)=S_1(0)P_1(0)=S_0P_0\cos(2\beta)\cos(arg(u))$ for the first
analyzer and
$T'(0)=S_1(0)P_1'(0)=S_0P_0\cos(2\beta)\cos(arg(u')+2\theta)$ for
the second. Both expressions have the same first factor which
arises because of the causal coupling between the photons. If we
don't use this causal coupling, (e.g. choose the angles in the
first two factors completely independent of each other), the
correlations vanish. The independent parameters $u$ and $u'$
represents local stochastic variables of this theory. Within the
distribution range, and depending on the angle $2\theta$, there
are certain values of $u'$ for which the second factor in $T'(0)$
can be negative, and certain values for which this factor is
positive. Because of these stochastic variables, we cannot know
which pair count combination (such as $(+-)$) we have until after
the coincident count is made. However, we still have distribution
information. The number of times the second factor in $T'(0)$ is
positive depends on the angle $2\theta$. As  a consequence, we
have correlations in the distributions which depend on the
relative angle. The presence of the local stochastic variables, a
necessary part of this theory, does not destroy all causal
correlation information.  We emphasize that this is a causal
non-local effect, and should not be confused with instantaneous
action at a distance effects. The correlations do not depend on
the distance between the analyzers, as long as the causal
polarization link between the photons remains intact until the
photons reach the analyzers.

 In comparison with the results
here, it is worth commenting on "Bell's theorem without
inequalities" type arguments that have appeared
\cite{Belinf,Greenb}. In studies leading to this work, the author
has calculated correlations with many different hidden variable
probability models, and not found one, no matter what sampling
distribution used, that agrees with the deterministic model and
quantum mechanics. This finding seems to be consistent with the
studies of hidden variable probability models by previous authors
\cite{Belinf,Greenb}. Replacing the deterministic transition
criterion at the analyzers with probability amounts to throwing
away much of the detailed correlations information.

\section{CONCLUSIONS}

We have introduced a quasi-deterministic analyzer model, and have
demonstrated with detailed calculations, that the law of Malus and
photon-photon correlations can be explained with a causal hidden
variable theory, via accumulation, one photon or photon pair at a
time. If we omit the causal coupling, the correlations are lost.
If we don't use a stochastic variable in the analyzer we can't
explain the data for the law of Malus nor the correlations. If we
replace the local deterministic decision process in the analyzers
with one based on probability, the correlations are reduced. In
simple terms, if we throw away the information, we can't explain
the data. Calculations to successfully explain some other quantum
measurements have been completed and will be reported on elsewhere
\cite{Dal2000}.

In this paper, we have only considered polarization correlations.
The essential features of this model are a causal link variable,
local analyzer variables, and a deterministic decision criterion.
Studies using a model with these features to describe energy-time
photon correlations \cite{Franzon,Tittle98,Tittle01} are underway.

\vskip 8.in

\vskip 1.5in
\vskip 1.5in
\vskip 1.5in
%
%


\begin{thebibliography}{}





\bibitem{Papal}
  Papaliolios C. (1967). Phys. Rev. Lett. 18, 622.
\bibitem{Belinf}
  Belinfante F. J. (1973). A Survey of Hidden Variable Theories,
  (Pergamon Press, New York).  See Ch. 5 for discussion of earlier
  studies  on the  law of Malus with hidden variable theories.
\bibitem{Ballen}
  Ballentine L. E. (1987). Am. J. Phys. 55 (9) 785. This is a
  resource article on this topic.
\bibitem{Afriat}
 Afriat A. and Selleri F. (1999). The Einstein Podolski and Rosen
 Paradox, (Plenum Press, New York). This recent book includes many
 references to earlier  studies.
\bibitem{Dal-G}
  Dalton B. J. (1997) in Causality and Locality in Modern Physics,
  Eds. G. Hunter, S. Jeffers, and J-P. Vigier, (Kluwer-Academic
  Publishers, Dordrecht, The Netherlands).
\bibitem{Strek}
 Strekalov D.V., Sergienko A. V., Klyshko D. N., and Shih Y. H., (1993).
 Phys. Rev. Lett, 74 (18) 3600.
\bibitem{Aspect-81}
 Aspect A., Grangier P. and Rogers G. (1981). Phys. Rev. Lett.  47, 460.
\bibitem{Aspect-82}
 Aspect A., Grangier P. and Rogers G., (1982). Phys. Rev. Lett.
 49,  91.
\bibitem{Alley}
 Alley C. O., Keiss T. E., Sergienko A. V. and Shih Y. H. (1994).
 in Frontiers of Fundamental Physics, eds.  M. Barren and F. Seller
 ( Plenum Press, New York).
\bibitem{Dal2000}
  Dalton Bill. Two particle correlations via quasi-deterministic
 analyzer model. Submitted for publication.
\bibitem{LR-M}
  Lamehi-Rachti M. and Mittig W. (1976).  Phys. rev. D 14 (10), 2543 .
\bibitem{Yariv}
 Yariv A. and Yeh P., (1984). Optical Waves in Crystals, (Wiley
 Interscience, New York).
\bibitem{McMas}
 McMaster W. H., (1954). Am. J. Phys. 22 (6) 351.
\bibitem{Fearn}
  Fearn H., James D. F. V. and Milonni P. W. (1996). Am. J. Phys.
 64, (18), 986.
\bibitem{Garucc}
 Garuccio A. and Rapisarda V. (1981).  Nuovo Cimento, 65A, 269.
 \bibitem{Bell}
  Bell, J. S. (1964) Physics 1, 195.
\bibitem{CHSH}
  Clauser J. F., Horne M. A., Shimony A., and Holt R. A., (1969) Phys. Rev. Lett.
  23, 880.
\bibitem{Greenb}
 Greenberg D. M., Horne M. A., Shimony A. and Zeilinger A. (1990).
 Am. J. Phys. 58, (12) 1131.
\bibitem{Franzon}
 Franzon, J. D. (1989) Phys. Rev. Lett. 62, 2205.
\bibitem{Tittle98}
 Tittle W., Brendel J., Zbinden H., and Gisin N., (1998)
  Phys. Rev. Lett 81, 3563.
\bibitem{Tittle01}
 Tittle W., Brendel J., Zbinden H., and Gisin N., (2001)
 Preprint from ArXiv:quant-ph/9806043.




\end{thebibliography}
\end{document}